**Mobility Analysis Workflow (MAW): An accessible, interoperable, and reproducible container system for processing raw mobile data**

Xiangyang Guan[a], Cynthia Chen[a,1], Ian Ren[b], Ka Yee Yeung[b], Ling-Hong Hung[b], Wes J. Lloyd[b]

[a] Department of Civil and Environmental Engineering, THINK lab, University of Washington, Seattle, WA 98195, USA, https://sites.uw.edu/thinklab
[b] School of Engineering and Technology, University of Washington, Tacoma, WA 98402, USA

[1]Corresponding author

**Abstract**

Mobility analysis, or understanding and modeling of people's mobility patterns in terms of when, where, and how people move from one place to another, is fundamentally important as such information is the basis for large-scale investment decisions on the nation's multi-modal transportation infrastructure. Recent rise of using passively generated mobile data from mobile devices have raised questions on using such data for capturing the mobility patterns of a population because: 1) there is a great variety of different kinds of mobile data and their respective properties are unknown; and 2) data pre-processing and analysis methods are often not explicitly reported. The high stakes involved with mobility analysis and issues associated with the passively generated mobile data call for mobility analysis (including data, methods and results) to be accessible to all, interoperable across different computing systems, reproducible and reusable by others. In this study, a container system named Mobility Analysis Workflow (MAW) that integrates data, methods and results, is developed. Built upon the containerization technology, MAW allows its users to easily create, configure, modify, execute and share their methods and results in the form of Docker containers. Tools for operationalizing MAW are also developed and made publicly available on GitHub. One use case of MAW is the comparative analysis for the impacts of different pre-processing and mobility analysis methods on inferred mobility patterns. This study finds that different pre-processing and analysis methods do have impacts on the resulting mobility patterns. The creation of MAW and a better understanding of the relationship between data, methods and resulting mobility patterns as facilitated by MAW represent an important first step toward promoting reproducibility and reusability in mobility analysis with passively-generated data.

**Highlights**

- A Mobility Analysis Workflow (MAW) is developed to process passively-generated data for mobility analysis.

- MAW facilitates accessibility, interoperability, reproducibility and reusability.

- MAW allows its components to be flexibly reused for a variety of mobility analysis.

- The study confirms the impacts of data pre-processing and analysis methods on the resulting mobility patterns.

# 1. Introduction

Mobility analysis, or understanding and modeling of people's mobility patterns in terms of when, where, and how people move from one place to another is fundamentally important. Such information is not only important for answering many scientific inquiries regarding how people interact with urban spaces and with each other, but also as a basis for many large- or mega-scale investment decisions on the nation's multi-modal transportation infrastructure. For decades, information on people's mobility patterns has been obtained from self-reported household travel surveys where randomly-selected respondents are asked to report all of their travel on one or two pre-determined travel survey days (Stopher and Greaves, 2007). Travel surveys, though providing rich information, are expensive (about $250-350 per household), and have relatively small sample sizes (typically ~0.1% of the region's population for urbanized areas). Because household travel surveys are conducted rather infrequently (once every few years), they are unsuitable for answering questions relating to how mobility patterns evolve over time or change after events.

The past two decades have seen a surge of studies using data from mobile devices to analyze individuals' mobility patterns (Chen et al., 2016). Such data often contains a large number of individuals (from hundreds of thousands to millions) while covering a sustained time period (from weeks to months and years). This data has two key pieces of information: the geographical locations (often expressed in longitude and latitude) where individual mobile devices are observed on the network, and the associated time when they are observed. Based on these two pieces of information, individuals' mobility patterns, in terms of when and where they go from one place to another, can be inferred.

Unlike travel surveys where trips are self-reported by the respondents and thus are automatically identified, data from mobile devices (hereafter "**mobile data**") needs to be pre-processed and analyzed to infer trips and their related information. This is due to the fact that mobile data is generated from users' opting into using certain mobile services (e.g., phone services, social media, or mobile apps) and consequently the amount and the quality of the mobile data vary greatly from one user to another depending on his/her usage patterns and the positioning technologies (e.g. GPS, WIFI, cellular tower, etc.) used by the data vendor. This means that pre-processing of the raw mobile data is first needed to correct data issues or to simply select a subset of users that meet certain criteria, followed by analysis of the pre-processed mobile data to infer mobility patterns. Those who use such data must go through these pre-processing and then analysis steps.

Exactly how this pre-processing and analysis sequence (called "**mobility analysis method**" hereafter for simplicity) is carried out is specific to individual researchers and indeed the proliferation of many studies that use big mobile data to infer mobility patterns suggest a variety of mobility analysis methods are being used (Chen et al., 2016). The different methods used can result in very different mobility patterns. Figure 1 illustrates an example. Both algorithms use a clustering method called trace segmentation in which a trajectory containing only raw mobile location records is first segmented into multiple subsequences, each satisfying the duration threshold. Then, the distance threshold is checked requiring that within a cluster of candidate records for an inferred stay, all pairs must be within the pre-set distance threshold. At the same distance and duration thresholds, one can see that the two algorithms result in quite different numbers of stays: at 0.5 minutes and 0.05 km, algorithm 1 results in 2 stays and algorithm 2 results in 11 stays; when the distance threshold increases to 0.5 km, the numbers become 3 and 9, respectively.

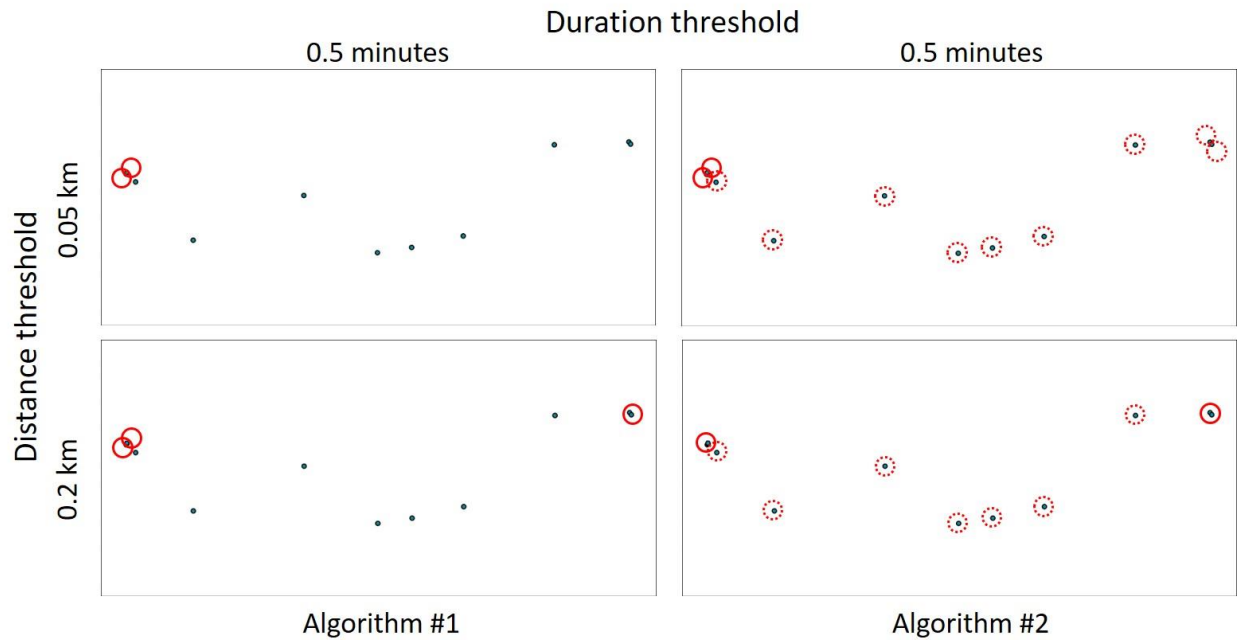

**Figure 1**. Illustration of two seemingly similar algorithms with the same distance and duration thresholds but completely different, resulting mobility patterns for a sample user with 19 records in the trajectory over two days. Algorithm #1 is the trace segmentation clustering algorithm used in this study. Algorithm #2 a similar trace segmentation clustering, implemented in the "preprocessing.detection.stops" function in the Python library "scikit-mobility (Pappalardo et al., 2021). Each black dot represents a raw location record; each solid red circle represents an inferred stay containing at least two location records; and each dashed red circle represents an inferred stay containing only one location record.

The sample illustrated in Figure 1 indicates that compared to the traditional survey data, the mobility patterns derived from those big datasets are much more sensitive to the actual pre-processing and analysis methods used: even essentially the same algorithms with the same distance and duration thresholds can result in vastly different patterns. Those resulting mobility patterns can directly impact policy making. This calls for a community effort involving everyone — not only those who have worked with mobile data, but also those who have not. This motivates the first aim of the current study, the development of a Mobility Analysis Workflow (**MAW**) that incorporates pre-processing and analyzing the raw mobile data to infer mobility patterns, which everyone can access and use with their own mobile data. More specifically, MAW supports four important properties of open science: accessibility, interoperability, reproducibility, and reusability. Accessibility requires not only the sharing of the code used for mobility analysis but also providing an easy-to-install and easy-to-configure version of the computer program or software with necessary dependencies, documentation, and licenses. Interoperability is defined as the ability to support information exchange between two or more systems (IEEE Standards Coordinating Committee, 1990). In the context of mobility analysis, this means a mobility analysis workflow can be tested and executed on different computing systems, including different operating systems (e.g. Mac OS vs. Windows), different hardware architectures (e.g. x86 vs. ARM), and different cloud platforms. Reproducibility is achieved if the same results are obtained when a mobility analysis workflow is tested with the same input data (Benureau and Rougier, 2018) by a different researcher or on a different computing system. Reusability implies that a workflow can be reused over additional cases other than those for which it was originally designed for (Lamprecht et al., 2020).

Even though no two mobility researchers use exactly the same methods to process and analyze mobile data, there are main commonalities and differences. In the pre-processing stage, as noted in Wang and Chen (2018), presence of oscillation is a primary issue in mobile data and not removing it can result in an overestimation of the regularity commonly found in human mobility patterns. Yet, there exist a variety of available methods and parameter settings in correcting oscillations, even for the same dataset. In the data analysis stage to derive the mobility patterns, even though clustering is a common method to identify stays (places where people perform activities), different clustering algorithms have been used (e.g., trace segmentation vs incremental clustering) with different threshold values (e.g., maximum distance or minimum duration required). Currently there is little to no systematic knowledge regarding how the differences in pre-processing and analysis of the mobile data impact the derived mobility patterns. Answering this question comprises the second aim of the study. To answer this question, six workflows that employ different mobility data pre-processing and analysis methods are designed to pre-process and analyze two commonly-used sample mobile datasets, and the resulting mobility patterns are compared among the six workflows. The findings provide additional evidence (on top of what is illustrated in Figure 1) in confirming the impacts of pre-processing and analysis methods on the derived mobility patterns.

The rest of the paper is organized as follows. Section 2 reviews how accessibility, interoperability, reproducibility, and reusability are accounted for in existing mobility research. Section 3 presents general design principles and features of MAW (detailed tutorials on how to use MAW can be found on GitHub (UW THINK lab github, 2021). Section 4 describes the six specific workflow designs used in this study to test and understand how different mobility analysis methods and their parameter settings affect the inferred mobility patterns. Results from those workflows are presented and analyzed in detail in Section 5. Section 6 summarizes this study and discusses potential future research directions.

## 2. State of the Art

### 2.1 Related work addressing accessibility, interoperability, reproducibility and reusability

The accessibility, interoperability, reproducibility and reusability of computer programs have been addressed through containerization. A container is a piece of software that includes code and dependencies that can be easily deployed. Docker is one of the most widely used technology for creating and managing containers (Docker Inc., 2013). It has been adopted in the research of bioengineering (Di Tommaso et al., 2017), information and communication technologies (Wan et al., 2018), social science (Kumar and Kurhekar, 2017), industrial engineering (Sollfrank et al., 2021), geology (Liu et al., 2020), and environmental science (Li, 2020). Containers enable users to run computer programs across different computing systems (i.e. interoperability) and produce the same results (i.e. reproducibility). Compared to virtual machines, they are also scalable (Merkel, 2014): a container packages up only the code and dependencies needed to run the code which include runtime, system tools, system libraries, and settings, and thus can be set up and deployed rapidly. Another factor that contributes to Docker container's popularity is its open-source nature (da Veiga Leprevost et al., 2017). There is a rich collection of containers in the public repositories of Docker Hub (https://hub.docker.com/). Those containers are not only for reuse but also for facilitating the creation of new containers; and users can also upload and publish their customized containers on Docker Hub (i.e. reusability). These features of Docker Hub greatly reduce the effort required to build a complex workflow and increase the accessibility and reusability of data science tasks (Madduri et al., 2019). To build workflows from containers, researchers have developed the Common Workflow Language (CWL) (Amstutz et al., 2015) and the Workflow Description Language (WDL) (OpenWDL, 2012) using existing data exchange formats such as JSON, YAML, or XML. These standards guide users to organize and connect multiple units of computation (e.g. containers) based on data processing logics. To apply these standards, it however requires knowledge about command-line tools which have complex syntax and are not user-friendly, especially to non-IT professionals.

Recent developments in creating accessible, interoperable, reproducible, and reusable workflows have focused on creating graphical user interfaces (GUIs) for containers and workflows. Two notable efforts in this line of work include the Galaxy platform (Afgan et al., 2018) and BioDepot-workflow-builder (Bwb) (Hung et al., 2019). The Galaxy platform is a web application that allows users to upload and analyze data. The limitation toward wider applications is that it only allows users to perform data analysis related to biomedical research. Bwb was developed as a portable and interactive graphical tool for creating modular workflows. It introduced widgets as the graphical interface for containers and represents workflows as connected widgets. Widgets make container/workflow building intuitive and interactive: users can import their own code or containers to create new widgets and existing widgets can be modified through a form-based user interface. However, as the Galaxy webserver, the Bwb has primarily served the biomedical research community.

In transportation research and practice, the use of accessible, interoperable, reproducible, and reusable workflows is still at a nascent stage. Existing efforts focus on the distribution of developed programs (typically in the form of code scripts) through platforms such as GitHub, JupyterLab or Google Colab (Boeing, 2020; Majka et al., 2019). As noted earlier, simply making scripts open-source does not address the dependency issues as aforementioned. To run the scripts, users need to figure out the compatible types of machines, operating systems, script compilers, and versions of packages and libraries. This information is not always clearly defined in a code repository, making it challenging to reproduce the research work and results. Though containers have recently attracted transportation researchers' attention (Feygin et al., 2020), they are used primarily to allow users to run models remotely, rather than for designing and implementing workflows.

## 2.2 Algorithms used in pre-processing and analysis of mobile data to derive mobility patterns

A common issue that is frequently encountered in mobile data is oscillation, which happens when a mobile device is seen jumping between different locations within a short time (within seconds) even though the device itself is not moving (Calabrese et al., 2011a, 2011b; Wu et al., 2014; Wang and Chen, 2018). The oscillation phenomenon generates a considerable number of records (~30%) that do not reflect devices' actual movements (Lee and Hou, 2006). A recent study (Wang and Chen, 2018) shows that not removing oscillated records prior to inferring mobility patterns over-stated the regularity property, which has been identified as the most striking feature of individual mobility patterns (González et al., 2008). Only a small portion of existing studies reported removing oscillations explicitly, primarily via heuristic methods (e.g., repeated visits to two or more locations within a short time, often a few seconds) (Wang and Chen, 2018). Clustering that is nearly applied by all studies to identify stays can remove oscillations to some extent, but Wang and Chen (2018) show that it is unsuitable for data generated from cellular towers (CDR data) or their triangulations (sightings data) due to sparsity issues.

To derive trips from big mobile data, the key is to identify stays from the data, i.e., instances where a user remains in the same place for some time to conduct activities such as staying at home, working, etc. After stays are identified, trips are simply movements from one stay (origin) to the next (destination). Methods to extract stays from mobile data are primarily of two kinds: 1) threshold-based methods; and 2) trace-segmentation methods. Threshold-based methods scan through the trajectories of each user (one trajectory refers to the user's available observations of one day) and identify stays by setting temporal and spatial thresholds such as spatial density, duration, speed and changes of heading (Hariharan and Toyama, 2004; Jiang et al., 2013). Threshold-based methods require thresholds to be predetermined, which are related to the characteristics of the data to be processed and are subject to analysts' knowledge on the data. Trace-segmentation methods (Zheng, 2015) first segment a trajectory into several sequences of consecutive observations and then identify a stay if a sequence of observations is bounded by both spatial and temporal constraints that correspond to the positioning error and the minimum time needed for conducting an activity. When applied to cellular data that is more sparse and less accurate (in terms of spatial uncertainty) than the GPS data (Chen et al., 2016), both of these two methods can be too strict in satisfying the temporal and spatial constraints (Wang and Chen, 2018). Thus, modification methods that first cluster observations close in space without considering the temporal information and then identify visits at each cluster have been proposed (Wang and Chen, 2018). This means that clustering methods may be applied multiple times on a user's trajectory to identify stays.

For the actual clustering methods used, threshold-based methods often use incremental clustering (Ester and Wittmann, 1998; Huang, 1998; Fisher, 1987), which treats the first location record in a person's trajectory as a cluster center, and then for each additional record, the algorithm calculates the distance between the cluster center and the new record. If the distance is below a pre-determined distance threshold, the cluster center is updated by incorporating the newly added record; otherwise, the original cluster center remains unchanged. Incremental clustering is sensitive to the order of the location records as inputs. To correct this order issue, K-means clustering has been proposed (Wang and Chen, 2018) to re-cluster those location centers initially identified from incremental clustering. The outputted final cluster centers from K-means clustering are inferred stays. The clustering method used in trace segmentation differs from incremental clustering in its distance calculation and the identification of the cluster centers. Incremental clustering calculates distances between a cluster center and a location point and then updates the center location every time a new record is identified as belonging to the cluster; trace segmentation, on the other hand, calculates all pairwise distances for all location records that satisfy a pre-determined duration threshold and then identifies the center of the cluster containing all records that also satisfy a pre-determined distance threshold. Computationally, incremental clustering is more efficient than trace segmentation. One commonality between the two clustering methods is that both require threshold values to be determined in advance. Different threshold values have been used. For example, the minimum stay duration ranges from 5 minutes for analyzing the call detail records data (Widhalm et al., 2015; Yin et al., 2018) to 10 minutes for cellular sighting data and GPS data (Bayir et al., 2009; Wan and Lin, 2013), and to 30 minutes for GPS, Bluetooth and Wi-Fi data (Ye et al. 2009; Vhaduri and Poellabauer, 2016). The distance thresholds used in those studies include 200 meters (Ye et al. 2009; Wan and Lin, 2013), 250 meters (Vhaduri and Poellabauer, 2016), 500 meters (Chin et al., 2019) and 1000 meters (Widhalm et al., 2015; Yin et al., 2018). As noted in the introduction section, there is currently no systematic knowledge on how the final resulting mobility patterns are affected by different factors including: the types of clustering algorithms used, various threshold values, whether to remove oscillation in the pre-processing of the data, and the order of clustering algorithms applied.

## 3. Development of Mobility Analysis Workflow (MAW)

### 3.1. Container Design

Individual containers are used to implement specific mobility analysis methods. Therefore, designing a container involves determining first which method a given container is to implement, and second the inputs, outputs, and the associated parameters. The container design process follows two principles: commonality and flexibility. Commonality means the containers in MAW should implement those methods that are the most frequently incorporated in mobility research. Flexibility refers to a container's capability of being reusable for a variety of mobility analysis workflows and this is done through a container's change points. Change points are key parameters used in an algorithm developed for a task. Values of the parameters can be changed by a user, thus called *change points*.

Based on the methods used in the current literature (see Section 2.2) and the commonality/flexibility considerations, MAW contains five containers: Trace Segmentation Clustering, Incremental Clustering, Stay Duration Calculator, Oscillation Corrector, and Stay

Integrator. Table 1 summarizes the functionalities of each container, inputs and outputs, and the associated change points.

3.2. Design of building blocks (modules) in the MAW

*3.2.1. Selection of containers in building modules*

The five containers can be used to design different workflows based on the context, data type, and purpose of the mobility analysis at hand. This section discusses a number of considerations relating to the design of the containers.

***Use of the Oscillation Corrector***. For mobile data in a low-density area, where oscillation is not prominent due to lack of high-rise buildings (which may block mobile device signals), either trace segmentation clustering or incremental clustering can be applied to the raw mobile data without calling for the Oscillation Corrector container. However, for cellular data or GPS data in a high-density area, the effect of oscillation can no longer be ignored and the container of Oscillation Corrector should be called when building a mobility analysis workflow. Whether the Oscillation Corrector container is applied to the raw mobile data or to clustering output (e.g., inferred stays) depends on the purpose of mobility analysis. For example, if a workflow is for real-time or online applications, deploying this container after clustering could be more desirable. This is because the number of location records in the raw mobile data is typically much larger than the number of clusters (stays). Therefore, applying the container to raw mobile data is likely to be much more computationally intensive than to clustering results from a previous step. On the other hand, in a non-time-sensitive situation such as retrospective analysis of people's mobility patterns, applying this container to raw mobile data before the clustering steps may be more suitable than applying afterwards, as it mitigates the effect of oscillation on clustering results.

**Table 1** Designs of the Five Containers in MAW

| Container name | Input and Output | Algorithm | Change point values |
|---|---|---|---|
| Trace Segmentation Clustering | *Input*: Location records[1] on a single day, sorted by time. *Output*: Same as inputs, with identified stay locations[2] added. Transient points[3] are denoted as -1. | It scans through all input data to identify subsequences of location records that satisfy the two thresholds (Hariharan and Toyama, 2004; Ye et al., 2009). First, the subsequence's duration (the time interval from the first to the last location record in the subsequence) must exceed a predefined duration[4] threshold. Second, any pairwise distance between two location records must fall below a predefined distance threshold. Those subsequences of location records are identified as stays, while location records not belonging to any subsequence are classified as transient points. | *Duration $\geq$ threshold* $\in$ [0.5, 30] minutes[5]. *Distance $\leq$ threshold* $\in$ [0.06, 0.25] km for GPS data[6], and [0.05, 1.0] km for cellular data[7]. |
| Incremental Clustering | *Input*: Location records (on either one day or multiple days). *Output*: Same as inputs, with identified stay locations added. Transient points are noted with placeholder of -1. | This container can be applied either to raw location records or inferred stays. When clustering location records, it starts by treating the first record as a cluster center, and then for each additional record, it calculates the distance between the cluster center and the new record (Fisher, 1987; Huang, 1998; Ester and Wittmann, 1998). If the distance is below a distance threshold, the cluster center is updated by incorporating the newly added record; otherwise, the location record forms a new cluster. Incremental clustering is sensitive to the order of the location records as inputs (Wang and Chen, 2018). Thus, K-means clustering (Kanungo et al., 2002) is applied to cluster location centers initially identified from incremental clustering. When applied to inferred stays, a duration threshold is first used to filter stays whose durations are longer than the threshold. Then the above procedures as to clustering location records are applied to the stay locations, by treating each stay location like a location record. | *Duration $\geq$ threshold* $\in$ [0.5, 30] minutes[5]. *Distance $\leq$ threshold* $\in$ [0.06, 0.25] km for GPS data[6], and [0.05, 1.0] km for cellular data[7]. |
| Stay Duration Calculator | *Input*: Location records with stay locations identified. *Output*: Same as input, with stay durations added. Transient points have a placeholder (e.g., -1) for stay durations. | It calculates the duration of each stay as the time interval between the first and the last of a sequence of location records associated with a stay. The change point—the minimum duration threshold—is to reassure that all stays satisfy the minimum duration requirement. Its value is typically set to be the same as the duration threshold in trace segmentation or incremental clustering. Any stay with a duration below the duration threshold will be removed and the location records associated with it will be outputted as transient points. | *Duration $\geq$ threshold* $\in$ [0.5, 30] minutes[5]. |

**Table 1** Designs of the Five Containers in MAW (cont'd)

| Container name | Input and Output | Algorithm | Change point values |
|---|---|---|---|
| Oscillation Corrector | *Input*: Location records with or without stay information, sorted by time. *Output*: Same as inputs with location records caused by oscillations removed. | Given a predetermined time window, it scans through the input data to identify subsequences of multiple stays or records that fall into the predetermined time window and contain at least one circular event (Wu et al., 2014). A circular event refers to a tour in which one device is initially found at one location, goes somewhere else and then returns to the previous location. For each unique location from all the subsequences suspected of oscillation, the total amount of time a person spends at each location across all days is calculated. The location where the person spends the longest time in a subsequence is treated as the true location (of a stay or location record) (Wang and Chen, 2018). Those in a subsequence but not regarded as the true visited location are deemed as from oscillation and their locations are updated with the one associated with the true location. | *Time window* $\in$ [1/6, 11] minutes[8]. |
| Stay Integrator | *Input*: Two sets of location records with inferred stay information attached, from two types of data (e.g. GPS and cellular), respectively. *Output*: Location records with stay information (stay location and duration) attached. | This container integrates stays of higher uncertainties (e.g. cellular stays) into stays of lower uncertainties (e.g. GPS stays). First, three temporal relationships between the two types of stays are defined (Peuquet, 1994): temporally separate, temporally contained and temporally intersecting. Additionally two spatial relationships are defined (Peuquet and Duan, 1995): spatially contiguous or not. Second, two stays of different types will either be merged, remain separate stays, or be split based on the above temporal-spatial relationships (Wang et al., 2019). Third, after integrating the stays, this container calls and executes four other containers, "Oscillation Corrector", "Stay Duration Calculator", "Incremental Clustering", and "Stay Duration Calculator" in a sequence. | *Duration* $\geq$ *threshold* $\in$ [0.5, 30] minutes[5]. *Distance* $\leq$ *threshold* = 0.2 km (Wang et al., 2019). |

[1] A location record is a data entry that records the location (in terms of geographical coordinates) of a mobile device, the timestamp when the location is recorded and the accuracy of the recorded location.

[2] Stay location is the geometric centroid of location records associated with the stay. In container design, a stay is identified as a sequence of location records.

[3] A transient point is a location record which indicates a passing-by location without stopping to carry out meaningful activities (e.g., a location record generated during navigation). In the container design, it is identified as a location record not associated with any stay.

[4] Duration of a stay is the time interval from the first to the last location record associated with the stay

[5] Based on Transportation Research Board (2005); Bayir et al. (2009); Ye et al. (2009); Gidófalvi and Dong (2012); Zhao et al. (2014); Vhaduri and Poellabauer (2016); Yin et al. (2018); Zhang et al. (2018).

[6] Based on Gidófalvi and Dong (2012); Montoliu et al. (2013); Vhaduri and Poellabauer (2018a, 2018b).

[7] Based on Montoliu et al. (2013); Vhaduri and Poellabauer (2018a, 2018b); Wang and Chen (2018); Yin et al. (2018); Zhao et al. (2018); Chin et al. (2019).

[8] Based on Tandon and Chan (2009); Shad et al. (2012); Fanourakis and Wac (2013); Wu et al. (2014); Qi et al. (2016); Katsikouli et al. (2019); Shan et al. (2019); Xu et al. (2021).

***Use of Trace Segmentation Clustering container vs Incremental Clustering container***.
There are three factors in choosing between the two clustering methods – trace segmentation and incremental clustering. The first factor is location accuracy. As noted in Section 3.1, distances in the two clustering methods are calculated differently: for trace segmentation clustering, pairwise distances between location records are calculated; and for incremental clustering, the distances between a location record and cluster centers are calculated. When the input location records have a large amount of noise such as in the cellular data, the distances calculated between them have large errors. Under this circumstance, calculating distances between location records and cluster centers can mitigate the magnitude of errors. Therefore, trace segmentation clustering is appropriate for high-accuracy location records (such as GPS data with location accuracy of only a few meters (Merry and Bettinger, 2019)) while for low-accuracy location records such as cellular data whose accuracy ranges from tens of meters to several kilometers (Järv et al., 2014), incremental clustering should be applied.

The second factor is temporal sparsity of the location records. For temporally sparse location records (i.e. large time intervals between consecutive location records), the use of trace segmentation clustering will result in that every segment (subsequence) only contains one or a few location records. This will make clustering infeasible or biased. Therefore, trace segmentation clustering is appropriate for location records with low temporal sparsity, and as temporal sparsity increases, incremental clustering should be considered.

The third factor is the purpose of clustering, especially when multiple days of location records are used. Trace segmentation clustering processes location records on a daily basis, while incremental clustering combines all days. Each has its pros and cons. Trace segmentation clustering can effectively identify transient points since it accounts for how long a person remains at one location, which is the key to distinguish between stays and transient points. However, it is also likely to misidentify a recurring stay location (e.g. home) to be different stay locations due to uncertainties in the stay locations. On the other hand, incremental clustering can capture recurring stays over multiple days, but it is also vulnerable to misidentifying recurring transient points as stays. For example, if a person passes by an intersection and leaves a location record every day, incremental clustering may falsely identify the intersection to be a stay location. Whether the purpose of clustering involves identifying recurring stays, transient points or both depends on understanding of the data generation process for the mobile data. As an example, the data generation process for some mobile data (an example is social media check-in data) may capture few transient points; and in such cases, incremental clustering can be applied alone.

It is worth noting that there are no hard criteria in selecting between the two clustering methods. In many cases the selection is a trade-off among multiple factors including the above three factors and others such as computational complexity. Also, in certain contexts, both clustering methods can be applicable. Examples of selecting the clustering methods and other MAW containers to construct workflows are given in Section 4.

## 4. Understanding the effects of pre-processing and analysis algorithms on inferred mobility patterns

To test how the inferred mobility patterns are affected by different pre-processing or analysis algorithms, their respective orders in a workflow, and change point values, two sample datasets are drawn from the real-world app-based data (see Section 4.1) and tested in different workflows. One dataset consists of predominantly GPS location records and the other of predominantly cellular location records. Two sets of metrics are evaluated, relating to the inferred mobility patterns and the computational costs.

## 4.1 Sample Selection

App-based data is multi-sourced data, generated from the use of various mobile phone applications (e.g., weather, shopping, dating, and navigation), each of which may use one or more positioning technologies, including GPS signaling or cellular tower triangulation. Each location record in the app-based data contains the device ID of an encrypted anonymous mobile device, a timestamp, a location (in the form of a pair of latitude and longitude coordinates), and the associated location accuracy in meters. A threshold of 100 meters is used to separate GPS data from cellular location records. If a location record has accuracy lower than 100 meters, this location record is assumed to be a GPS location record; and otherwise, it is a cellular location record. The data used in this study is the GPS and cellular data generated when people use mobile phone apps (thus called app-based data) in the central Puget Sound region covering the four counties (King, Kitsap, Pierce, and Snohomish) during the period of March, April, and November of 2019. We then randomly selected 1,000 users whose location records contain over 80% GPS location records. This dataset constitutes our sample GPS dataset. Similarly, a random set of 1000 users each with more than 80% of location records as cellular location records are selected, which comprises the sample cellular dataset.

The sample GPS dataset and cellular dataset contain 1,880,818 and 192,441 location records, respectively. The number of location records per person varies over time, as shown in Figure 2. The daily number of location records per person in the GPS data varies between 82 and 183 during the months of March and April, and drops to around 50 in November 2019. Meanwhile, the daily number of location records per person in the cellular dataset remains around 40 throughout the study period.

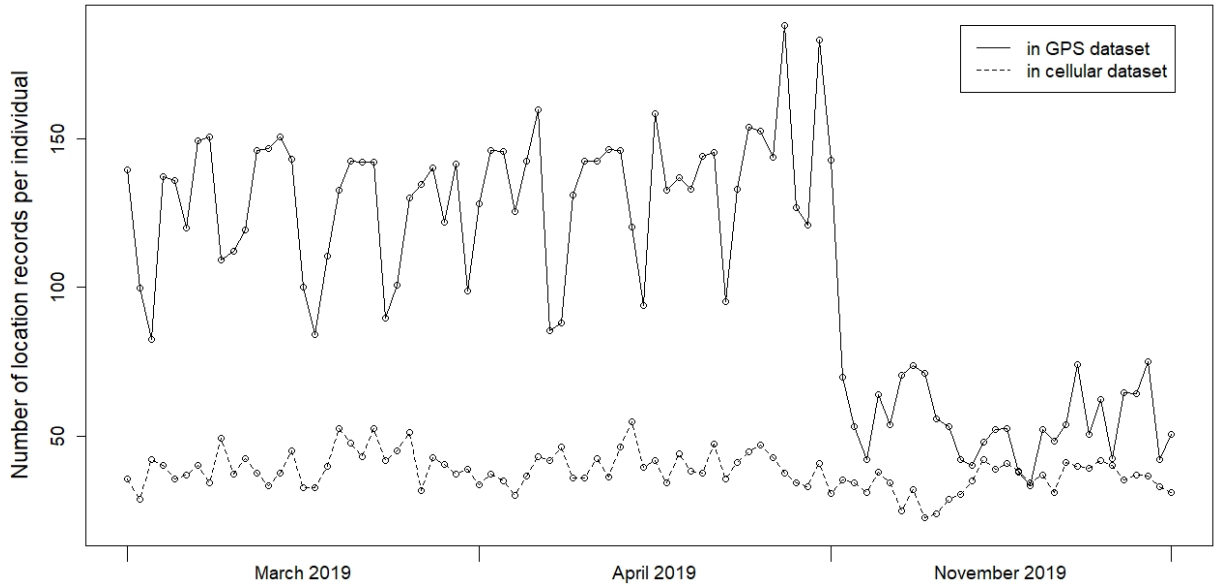

**Figure 2**. Number of location records per person every day over the study period, in the GPS data and cellular data, respectively.

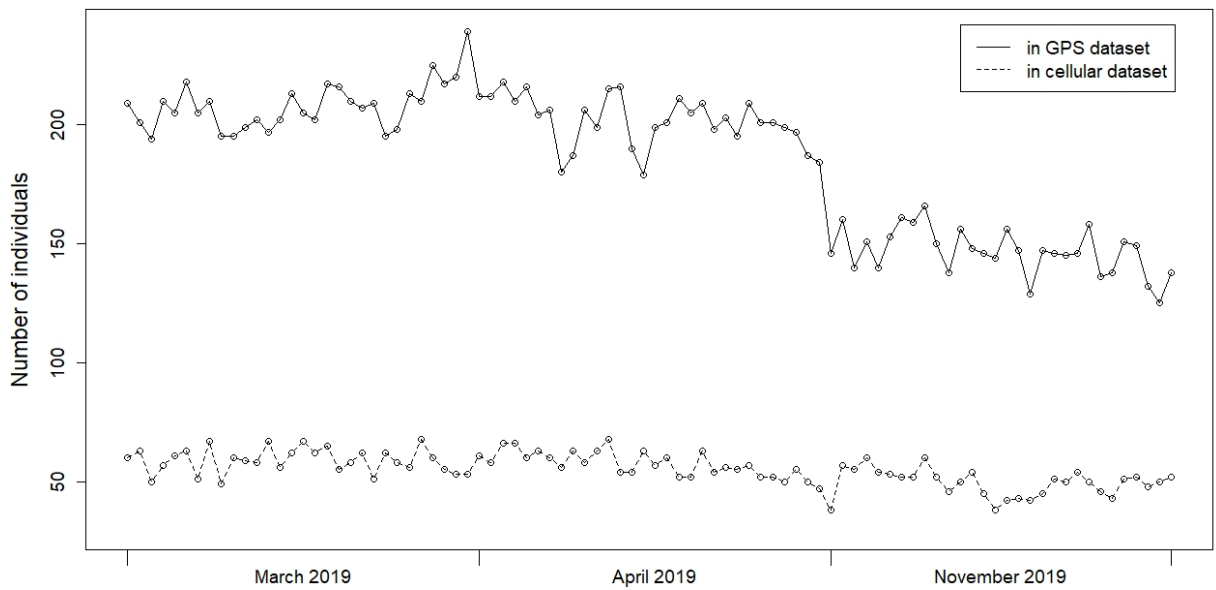

**Figure 3**. Number of people observed every day over the study period, in the GPS data and cellular data, respectively.

The number of people observed in the datasets each day has similar trends, as shown in Figure 3. Around 200 people are observed each day in the GPS dataset during March and April, and the number drops to around 150 in November. In the cellular dataset, the number of people observed every day remains around 50.

To understand the sparsity of the data, two quantitative measures are investigated. The first one is for how many days each person is observed, and Figure 4 shows the distribution of this number for the GPS and cellular datasets, respectively. In the GPS dataset, 19% of the 1000 people are observed for only 1 day, while in the cellular dataset, this percentage increases sharply to 62%. On the other hand, while both datasets see a few people being observed every day during the study period (2 people in the GPS dataset and 1 person in the cellular dataset), the top 10% quantile with the highest number of days observed has a minimum of 55 days observed in the GPS dataset and only 13 days observed in the cellular dataset. Therefore, people in the cellular dataset are observed overall for fewer days than those in the GPS dataset. Second, the cellular location records have longer observation time intervals than the GPS location records, as shown in Figure 5. Observation time interval measure the time (in minutes) between consecutively observed location records. In the GPS dataset, 9.3% of the location records have an observation interval less than 10 minutes, while the cellular dataset has only 6.6%. While in both datasets, the observation interval can get as large as 23 hours, the median observation intervals in the GPS and cellular datasets are 0.1 minute (i.e. 6 seconds) and 6 minutes, respectively. These findings suggest the cellular data is considerably sparser than the GPS data.

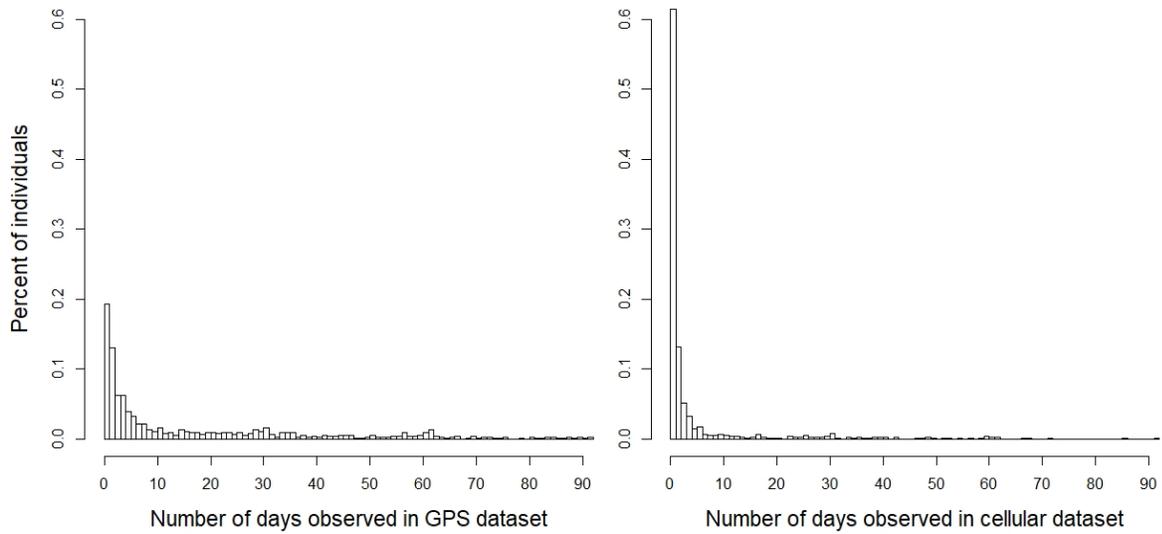

**Figure 4**. Distribution in the number of days observed, in the GPS dataset and cellular dataset, respectively.

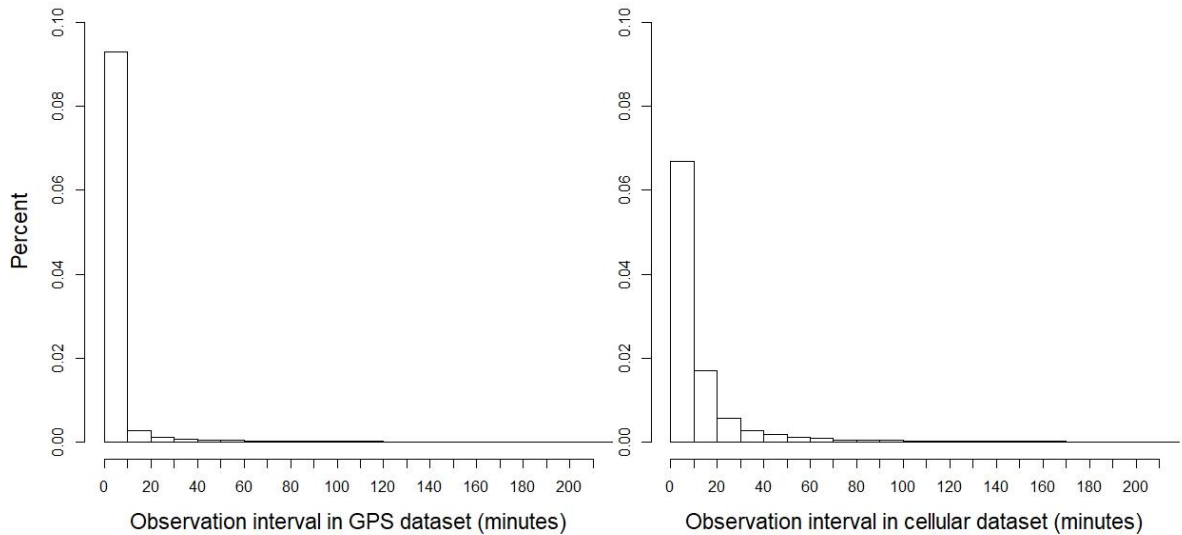

**Figure 5**. Distribution in the observation interval (time between consecutive location records), in the GPS dataset and cellular dataset, respectively. The two subfigures show only the distributions for observation interval less than 210 minutes, as more than 99% of the location records in both datasets have observation interval below 210 minutes.

## 4.2 Workflow Designs

### 4.2.1 Relating to the Oscillation Corrector container

A set of workflows are designed to test how the inferred mobility patterns are affected by whether oscillation is addressed, when to address it, and its change point value settings. Considering that (1) oscillation is a prominent issue in cellular data (Qi et al., 2016); and (2) only incremental clustering is appropriate for analyzing cellular data due to its low location accuracy and high spatial-temporal sparsity (see Section 3.2), three workflows are to be tested (Table 2).

**Table 2** Workflows for testing different settings of the Oscillation Corrector container

| Workflow index | Workflow design | Note | Change point values |
|---|---|---|---|
| 1 | Incremental Clustering – Stay Duration Calculator | Oscillation is not dealt with. | Distance ≤ threshold = 1 km[1]; Duration ≥ threshold = 5 minutes[1]; Time window ∈ {1/6, 5, 11} minutes. |
| 2 | Incremental Clustering – Stay Duration Calculator – Oscillation Corrector – Stay Duration Calculator | Oscillation is addressed as a post-processing step. | |
| 3 | Oscillation Corrector – Incremental Clustering – Stay Duration Calculator | Oscillation is addressed as a pre-processing step. | |

[1] These values were justified in Wang and Chen (2018), which used cellular data collected in the same region as the sample cellular data.

**Workflows 1 through 3** address the situations where oscillation is simply ignored, dealt with in post-processing (i.e. after inferring stays), and corrected in pre-processing (i.e. before inferring stays), respectively. Different change point values of Oscillation Corrector are tested in workflows that include this container. In both workflows 1 and 2, incremental clustering is applied directly to the raw cellular data, followed by one step to calculate the durations of inferred stays. Workflow 2 additionally post-processes the inferred stays by removing those suspected for oscillation. Since this operation may change the cluster compositions (e.g. two clusters merged into one), a Stay Duration Calculator container is subsequently called to update the stay durations. Workflow 3 in contrast, addresses oscillation beforehand, prior to applying incremental clustering and calculating stay duration.

For change point values, the distance threshold and duration threshold are set to be the ones used by Wang and Chen (2018), which are appropriate for the sample cellular data. Different values for the time window are tested to understand how they affect the performance of Oscillation Corrector. In previous studies, the time window for detecting oscillations has been set to 10-20 seconds (Qi et al., 2016), 1 minute (Katsikouli et al., 2019), 2 minutes (Wu et al., 2014), 5 minutes (Xu et al., 2021), and 11 minutes (Shad et al., 2012). Three values, which include 10 seconds (the lower bond), 11 minutes (the upper bound) and 5 minutes (the middle point) are thus tested for the time window in Oscillation Corrector.

*4.2.2 Relating to different clustering methods*

Trace segmentation clustering and incremental clustering are the two most widely used clustering methods for analyzing mobile data to infer stays. Therefore, a set of workflows are designed to test how using different clustering methods leads to different inferred mobility patterns. For this group of workflows, the sample GPS dataset is used as the input data, as both clustering methods are applicable to GPS data (if using the sample cellular dataset, only incremental clustering would be applicable). The workflows to be tested are listed in Table 3.

**Table 3** Workflows for testing different clustering methods

| *Workflow index* | *Workflow design* | *Note* | *Change point values* |
|---|---|---|---|
| 4 | Incremental Clustering – Stay Duration Calculator | Only incremental clustering is applied. | Distance threshold ∈ {0.05, 0.2, 0.5} km; Duration threshold ∈ {0.5, 5, 30} minutes |
| 5 | Trace Segmentation Clustering – Stay Duration Calculator | Only trace segmentation clustering is applied. | |
| 6 | Trace Segmentation Clustering – Incremental Clustering – Stay Duration Calculator | Both trace segmentation clustering and incremental clustering are applied. | |

**Workflows 4, 5** and **6** concern the situations where only incremental clustering, only trace segmentation clustering, and both trace segmentation clustering and incremental clustering are applied to the GPS data, respectively. In each workflow, a Stay Duration Calculator container follows the clustering step(s) to calculate the durations of inferred stays. For workflow 6, the particular order in which trace segmentation clustering comes ahead of incremental clustering is based on the consideration that when the two clustering methods are applied together, the purpose of incremental clustering is to identify recurring stays over multiple days (see Section 3.2). Therefore, incremental clustering needs preliminary stays inferred by trace segmentation clustering as its inputs.

In terms of change point vales, various options as discussed in Section 3.1 are tested. There are two change points in these three workflows, the distance threshold in the clustering algorithms and duration threshold in the clustering algorithms and Stay Duration Calculator container. In existing literature, values adopted for the duration threshold include 30 seconds (Gidófalvi and Dong, 2012), 5 minutes (Yin et al., 2018), 10 minutes (Zhao et al., 2014; Vhaduri and Poellabauer, 2016), and 30 minutes (Ye et al., 2009; Zhang et al., 2018). For the distance threshold, values of between 0.05 and 0.1 km (Yin et al., 2018), between 0.06 and 0.25 km (Vhaduri and Poellabauer, 2018a), 0.15 km (Gidófalvi and Dong, 2012), 0.25 km (Montoliu et al., 2013; Vhaduri and Poellabauer, 2018b), and 0.5 km (Zhao et al., 2018) have been observed in existing literature. The lower bound, upper bound and a middle point of each change point are selected for testing, which results in distance threshold $\in \{0.05, 0.2, 0.5\}$ km and duration threshold $\in \{0.5, 5, 30\}$ minutes.

## 4.3 Performance metrics

Two categories of performance metrics are adopted to quantify the effects of pre-processing and analyzing algorithms on inferred mobility patterns. The first category refers to the spatial-temporal patterns of human mobility (Chen et al., 2016), and includes the following specific performance metrics:

- Number of trips per person per day.
- Radius of gyration per person per day: a measure of the activity space of a person on a daily basis. The daily radius of gyration is calculated using equation 1 (Lu et al., 2012),

$$r_g(i) = \sqrt{\frac{1}{N}\sum_a [\text{dist}(s_a, \bar{s})]^2} \qquad (1)$$

  where index *i* refers to a person, index *a* refers to the person's stays, $s_a$ represents one stay, $\bar{s}$ is the center of mass for all stays of the person, $\text{dist}$ denotes the (Euclidean) distance between two stay locations, and *N* is the number of stays made by the person. For each day, $r_g(i)$ is calculated using stays on that day. The calculated $r_g(i)$ is then averaged over all days that have at least one stay.

- Departure time distribution. The departure time of a trip is assumed to be the time when the last stay ends. To quantify the departure time distribution, every day is divided into 48 half-hour intervals, and the number of trips starting at each interval is counted. The departure time distribution summarizes the departure time of all trips made by all people on all days.

The second category of performance metrics relates to the computational performance of workflows. The key metrics in this category are the computation time of a given workflow, measured as the time interval from when the workflow begins reading input data to when it finishes printing the output, and the memory usage over time throughout the running of the workflow.

## 5. Results

5.1. Effect of correcting oscillation on inferred mobility patterns

Figure 6 shows how the inferred number of trips and radius of gyration change when the Oscillation Corrector is applied differently in a workflow. It is worth noting that based on the definition of radius of gyration (Section 4.3), a person must have at least one stay for the radius of gyration to be meaningful. Therefore, all results on the number of trips and radius of gyration are reported only for users with at least one stay inferred in the three-month period by all workflows designed for the corresponding dataset. For the cellular dataset, 483 out of the 1,000 users meet this requirement. And the number for the GPS dataset is 659.

There are two observations from Figure 6. The first observation is that when oscillation corrector is applied as pre-processing vs. as post-processing, its effects on the inferred mobility patterns are different. For the number of trips, correcting oscillation after clustering decreases the number of trips inferred, from 1.46 when Oscillation Corrector is not applied to 1.34 when Oscillation Corrector is applied as post-processing with an 11-minute time window. This finding is reasonable, because Oscillation Corrector as a post-processing step will take inferred stays (from the clustering step) as input, and when an oscillation is identified, the stay caused by the oscillation will be merged with the true stay. This leads to a decrease in the number of stays and consequently decreased number of trips after applying oscillation corrector.

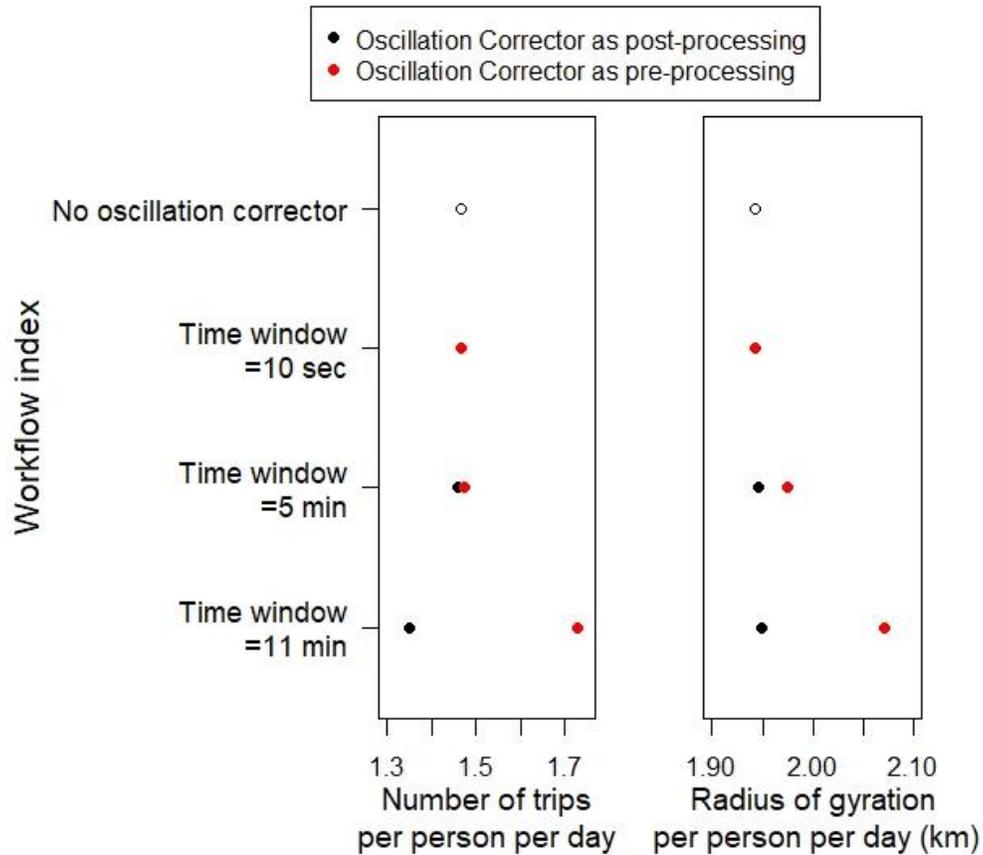

**Figure 6**. Effects of Oscillation Corrector on number of trips and radius of gyration. Both metrics are calculated on a daily basis for each person, and average over all days and all people.

What maybe a bit surprising is that when applied before clustering (i.e. as pre-processing), Oscillation Corrector increases the number of trips. This finding is consistent with existing literature (Xu et al., 2021). To further understand the mechanism behind this increase, a user from the cellular dataset is randomly drawn, and his/her trajectories before and after applying Oscillation Corrector as pre-processing are shown in Figure 7.

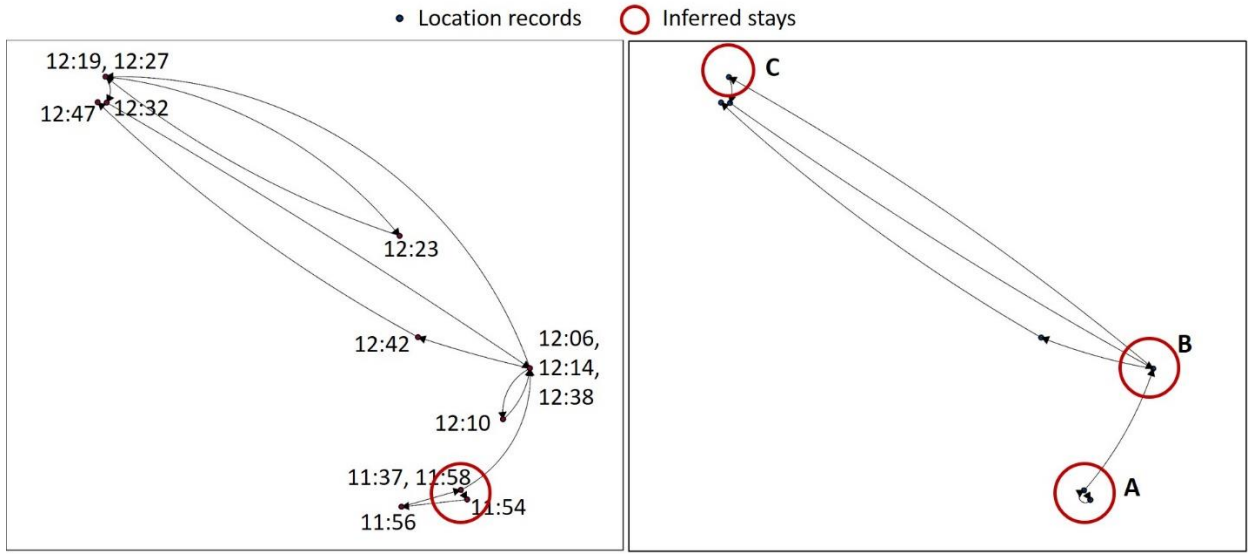

**Figure 7**. Sample user trajectory to explain the increase in the number of trips after applying Oscillation Corrector as pre-processing. The left figure shows the trajectory without correcting oscillation, and the right figure shows the trajectory after applying Oscillation Corrector in pre-processing. Black dots represent raw location records and red circles represent inferred stays.

The user shown in Figure 7 has three stays (right figure), two of which cannot be inferred when oscillation is present (left figure). The reason is that oscillation breaks each of the two stays into two segments, and neither segment satisfies the duration constraint (i.e. a minimum duration of 5 minutes). Stay B (on right figure), for example, lasts from 12:06 to 12:14, which meets the duration constraint. However, an oscillation occurring at 12:10 separates the stay into two parts, one from 12:06 to 12:10 and the other from 12:10 to 12:14, and both parts are below the duration constraint of 5 minutes (shown in the left figure). Similarly for stay C, an oscillation happening at 12:23 breaks the stay (lasting from 12:19 to 12:27) into one segment lasting from 12:19 to 12:23 and the other from 12:23 to 12:27, respectively. The fact that presence of oscillation breaks a user's trajectory into segments of short durations (less than the minimum duration constraint) and thus makes the clustering method unable to identify those stays likely underlies the observation that correcting the oscillations before clustering will increase the number of stays and trips. Among the 483 users who made at least one stay in the cellular dataset, 101 users (21%) have increased number of stays/trips observed after correcting oscillations in pre-processing, suggesting this pattern is common in a cellular dataset. It is also worth noting that stays with shorter durations are more vulnerable to the impact of oscillation than stays with longer durations: in Figure 6, oscillation also happened to stay A but does not mask stay A from being inferred, likely due to the longer duration of stay A (19 minutes) than B and C (8 minutes).

For radius of gyration, Figure 6 suggests two patterns. First, correcting oscillation will increase the radius of gyration per person per day. The value increases slightly from 1.943 km when oscillation is not corrected to 1.946 km and 1.949 km when Oscillation Corrector is applied as post-processing with time windows of 5 min and 11 min, respectively. When Oscillation Corrector is applied as pre-processing under the two time windows, the radius of gyration increases to 1.975 km and 2.071 km, respectively. These numbers lead to the second observation, that is, with the same change point value, applying Oscillation Corrector as pre-processing results in larger radius of gyration than applying it as post-processing. As aforementioned, Oscillation Corrector as pre-processing could lead additional stays to be identified (thus increases the number of stays) while Oscillation Corrector as post-processing could result in stays being merged (thus decreases the number of stays). It is intuitive that those merged stays are close in distance while those newly identified stays are likely far away from existing stays (or otherwise they may be absorbed into existing stays). Therefore, both the increased number of stays by Oscillation Corrector as pre-processing and decreased number of stays by Oscillation Corrector as post-processing reduce the spatial concentration of stays and lead to increases in radius of gyration.

Another mobility metric of interest for comparison is departure time distribution. Figure 8 shows the departure time distribution in terms of the proportion of trips that start at different times of day, aggregated to half-hour intervals, under different configurations for Oscillation Corrector. When the time window equals 10 seconds or 5 minutes, there is little to no difference in the departure time distributions between not correcting oscillation, applying Oscillation Corrector in post-processing or in pre-processing. However, as the time window further increases to 11 minutes, the differences among the three configurations become more noticeable. There are two observations under the situation where the time window is equal to 11 minutes in Figure 8. First, applying Oscillation Corrector as pre-processing appears to result in more trips happening in early morning (12am – 6am) and fewer trips during the morning peak (8am – 11am) and afternoon peak (3pm – 6pm), compared to the pattern for not correcting oscillations; and second, correcting oscillations in post-processing appears to have opposite effects: it results in fewer trips inferred in early morning and more trips inferred during the morning and afternoon peaks. This finding may be related to the earlier finding that Oscillation Corrector increases the number of trips if applied in pre-processing and decreases the number of trips if applied in post-processing (Figure 6). It may also be related to the fact that in early morning, when cellular phone usage is the lowest during a day (Caceres et al., 2012) and thus data during that period is fairly sparse. The more sparse the data is, the fewer location records that capture each stay, and the more likely that stay inference will be affected by oscillations (and correcting oscillations). As a measure of sparsity (Ban et al., 2018) in the cellular dataset, the average time interval between consecutive location records is 1,442 seconds in early morning while only 854 during other times of day, suggesting that the cellular data is more sparse in early morning than during other times of day.

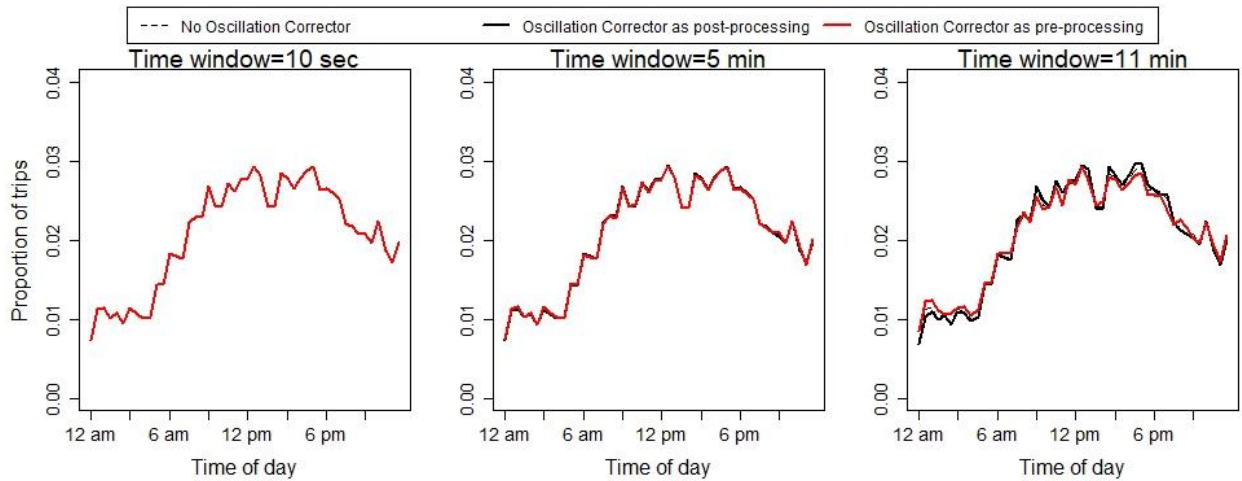

**Figure 8**. Effects of different settings for Oscillation Corrector on departure time distribution. The black dash curve and black solid curve mostly overlap in the figure.

5.2. Effect of different clustering methods on inferred mobility patterns

Two clustering methods, trace segmentation clustering and incremental clustering, with varying change point values, are tested in this study. Figure 9 shows how they result in different numbers of trips inferred. The first observation from Figure 9 is that when trace segmentation clustering is applied, the number of trips inferred is always higher than when (only) incremental clustering is applied, and this phenomenon is more pronounced when the distance threshold is lower. There are two possible reasons behind this phenomenon. The first one is associated with how the distance threshold is applied differently in trace segmentation clustering and incremental clustering. In trace segmentation clustering, the distance between every pair of location records must be below the distance threshold for the location records to be clustered, compared to the distance between cluster center and a location record considered in incremental clustering. This increases the difficulty for location records to be clustered together in trace segmentation clustering, and thus make them more likely to break into different small clusters than in incremental clustering. The second reason is because trace segmentation clustering divides a user's trajectory into segments and identifies stays in each segment. There is likelihood that one stay be identified as two or more stays. The segmentation operation also explains why increasing distance threshold raises the number of trips inferred by incremental clustering more than the number inferred by trace segmentation clustering: the (temporal) segments created by trace segmentation clustering pose an additional constraint for the inferred number of trips to increase, while incremental clustering is free from this constraint. The second observation from Figure 8 is that applying incremental clustering after trace segmentation clustering makes little difference compared to using trace segmentation clustering only. This finding is expected as incremental clustering applied after trace segmentation clustering only changes the labeling of stays, i.e. recurring stays will be identified to be at the same stay location. Its effect on the number of stays (and thus trips) is minimum.

Regarding how the change point values affect the inferred number of trips, there are also two observations. The first one is that the number of trips decreases with duration threshold, under each given distance threshold. This observation is expected, as higher duration threshold results in more records to be clustered together, resulting in fewer stays. The second observation is that when the duration threshold is low (e.g., at 0.5 minute), the number of trips can increase with higher distance threshold. Figure 1 in the introduction section illustrates this phenomenon. At the distance threshold is very low (0.05 km), fewer records satisfy this threshold, resulting in few stays. As the distance threshold increases to 0.2 km, more stays emerge as more records can satisfy this criterion. When the duration threshold is at 5 minutes or higher, this distance effect disappears: Figure 9 shows that the number of trips remains roughly the same at varying distance thresholds.

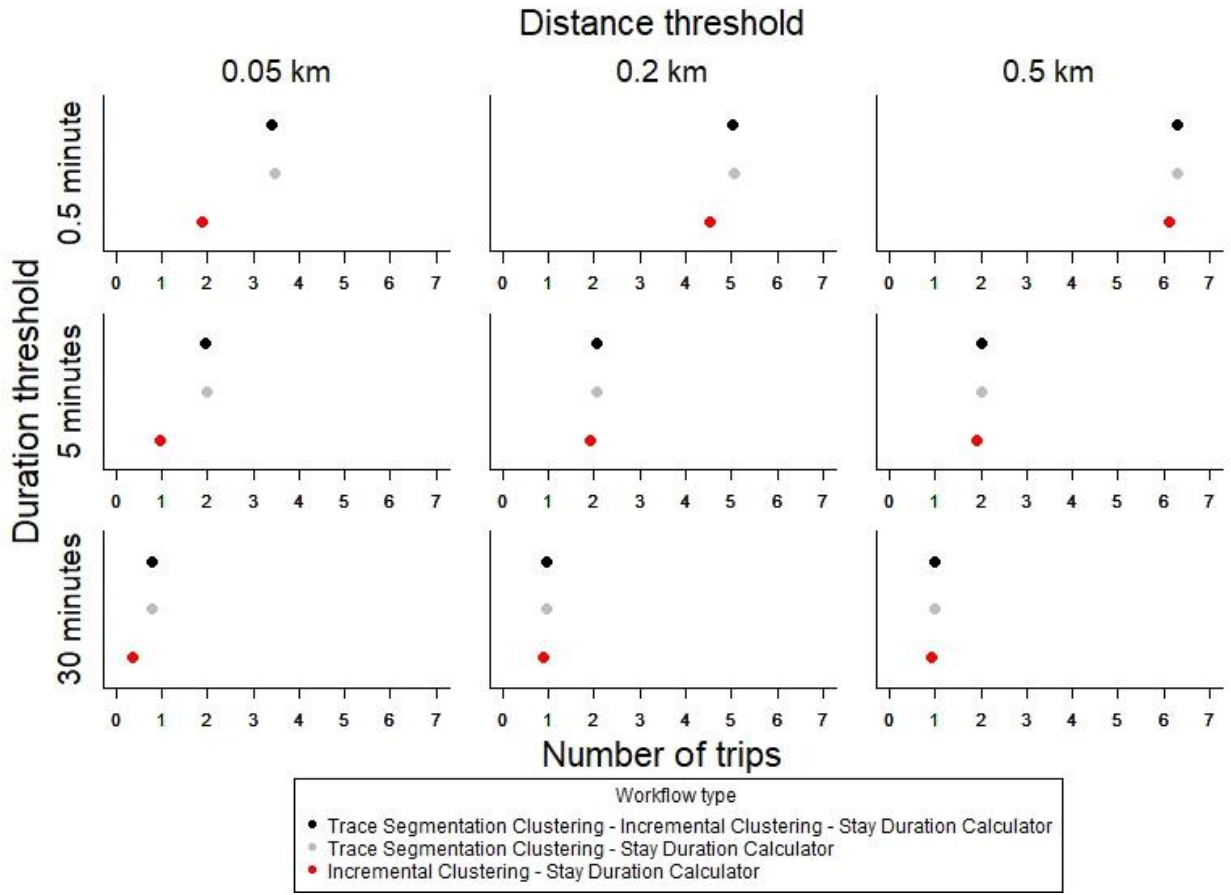

**Figure 9**. Effects of different types of clustering methods and different change point values on the inferred number of trips.

How different settings of clustering methods affect radius of gyration (per person per day) is shown in Figure 10. We make three observations here. First, increasing duration threshold leads to lower radius of gyration, with everything else being equal. Second, the radius of gyration resulting from incremental clustering appears to be always smaller than that resulting from trace segmentation clustering. Both findings are consistent with previous findings and expected. What is slightly different from Figure 9 on the number of trips is that the increase in radius of gyration with distance is more pronounced at even higher duration thresholds such as 30 minutes.

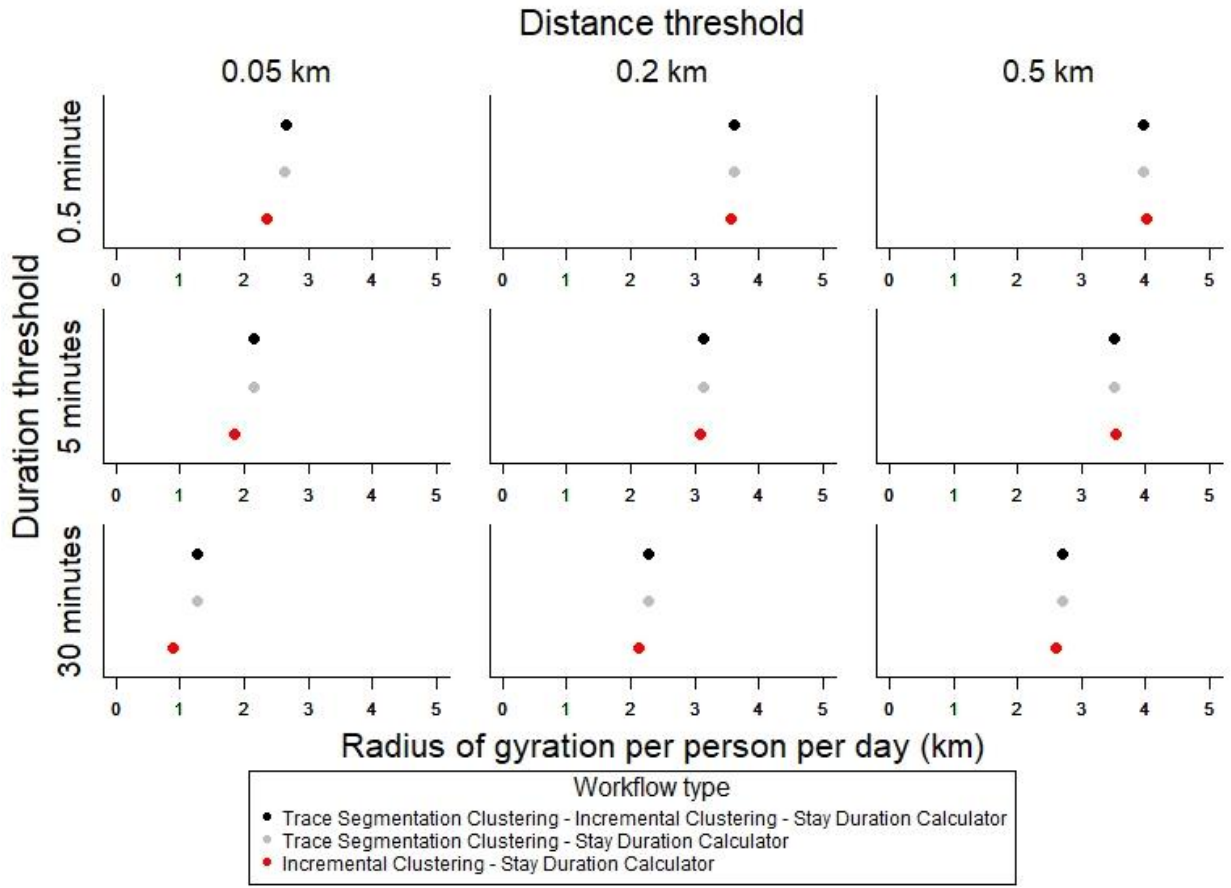

**Figure 10**. Effects of different types of clustering methods and different change point values on the inferred radius of gyration, averaged for all days for all people.

Figure 11 shows the departure time distribution under different clustering method configurations. Three observations are as following. First, as distance threshold increases, the distribution introduces more variations, meaning that peaks are higher and valleys are lower. The same goes for decreasing duration threshold. The reason behind these two observations could be that as the constraints are relaxed (raising distance threshold or lowering duration threshold), the inferred stays will falsely capture more and more transient points when people are traveling (e.g. waiting for green light at an intersection). For example, at distance threshold of 0.5 km and duration threshold of 0.5 minutes, the departure time distribution seems to indicate a morning peak around 8 am and evening peak around 6 pm.

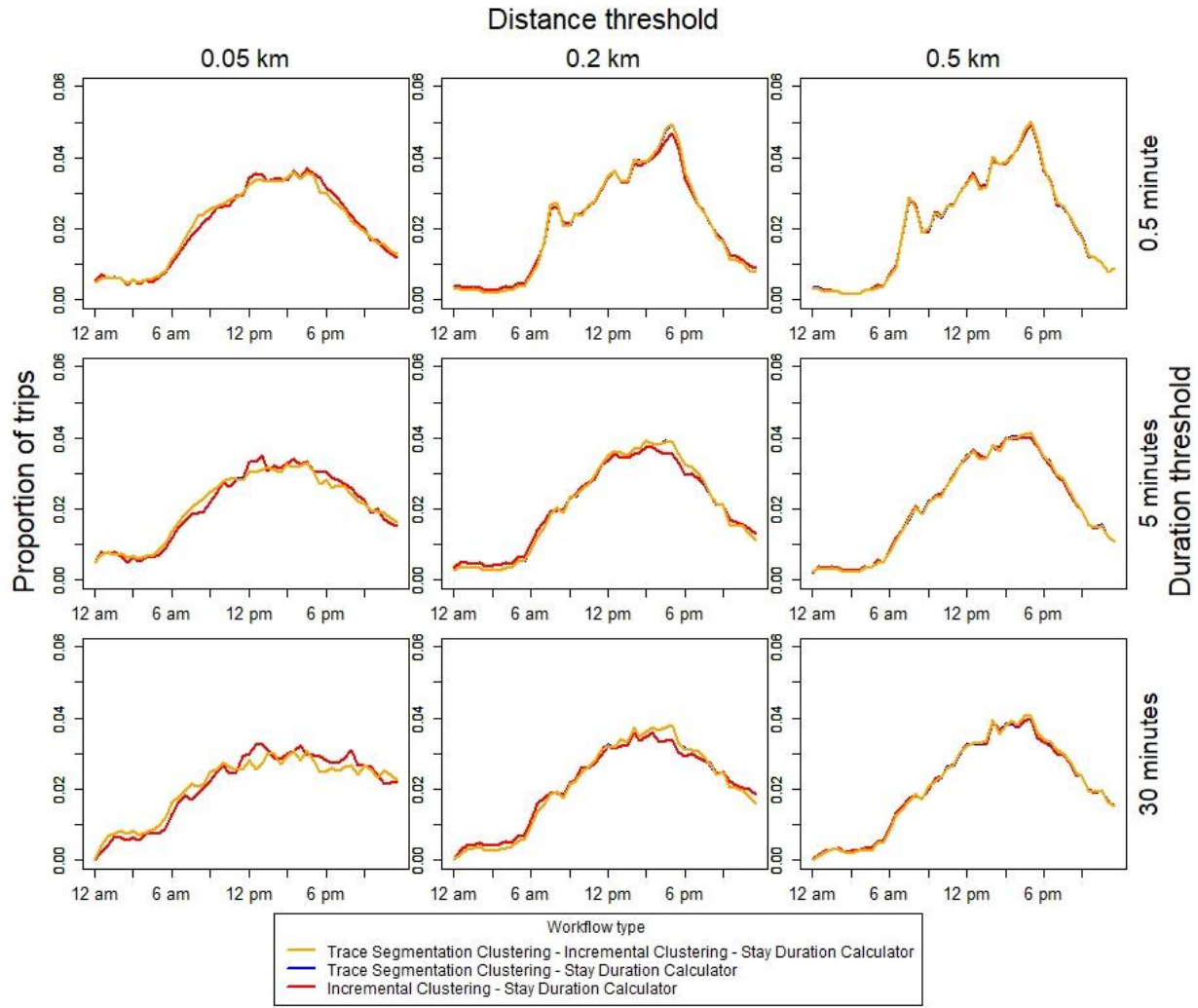

**Figure 11**. Effects of different types of clustering methods and different change point values on the inferred departure time distribution. The orange curve and blue curve mostly overlap in the figure.

For comparison between incremental clustering and trace segmentation clustering, Figure 11 suggests that the departure distributions inferred by the two clustering methods are overall similar, especially when distance threshold is high and duration threshold is low. However, a noticeable difference between the departure time distributions inferred by the two clustering methods is that when the distance threshold is 0.05 km, a smaller proportion of people travel at night (9 pm to 9 am) as inferred by incremental clustering, compared to that inferred by trace segmentation clustering. But when the distance threshold is 0.2 km, the pattern reverses. Considering that most people stay at home at night and those stays are recurring, incremental clustering, which utilizes location records of multiple days and applies the distance threshold to the distance between a cluster center and a location record, could capture home stays more accurately than trace segmentation clustering. Trace segmentation clustering on the other hand, could falsely identify a home stay to be multiple separate stays especially when the distance threshold is low (e.g. 0.05 km). However, incremental clustering has its disadvantages as well. When multiple stays form a cluster with close distances between them, incremental clustering could falsely merge them into a single stay. This may explain why incremental clustering infers fewer trips during the day (9 am to 9 pm) than trace segmentation clustering when the distance threshold equals 0.2 km.

## 5.3. Computational performance of the workflows

To generate the datasets for testing the computational performance of workflows, users from the app-based data (see Section 4.1) are randomly sampled. All location records of a sampled user are added to a dataset until the dataset reaches a given size. Using this procedure, six datasets at 1 GB, 10 GB, 25 GB, 50 GB, 75 GB, and 100 GB, respectively, are generated.

The workflow applied to each dataset is as following. First, the dataset is split into two, one containing only cellular location records and the other containing only GPS location records (method of splitting data is described in Section 4.1). Then the workflow "Trace Segmentation Clustering – Incremental Clustering – Stay Duration Calculator" is applied to the GPS dataset, and the workflow "Incremental Clustering – Stay Duration Calculator – Oscillation Corrector – Stay Duration Calculator" is applied to the cellular dataset. The outputs from the two workflows (i.e. inferred GSP and cellular stays) then serve as inputs to the Stay Integrator container to produce the final output stays.

Due to the large data volume, the tests are performed using University of Washington's high-performance computing cluster "Hyak". All tests are performed on a Hyak computing node that has an Intel Xeon Gold 6230 processor with 20 CPU cores each at 2.10 GHz, and 64 GB memory.

Figure 12 shows the elapsed computation time and memory utilization when the 10 GB app-based dataset (which contains both GPS and cellular location records) is processed. The horizontal axis represents the time elapsed in minutes, and the vertical axis shows the corresponding memory utilization over time. The red dashed lines indicate when the execution of a container is completed. For example, the Trace Segmentation Clustering container for processing the GPS dataset (10 GB) was finished in 58 minutes after the workflow starts. The lowest memory utilization on average appeared between 121 minutes and 124 minutes when the Oscillation Corrector container is applied to cellular stays, while the highest memory utilization happens when the Incremental Clustering container is running to process the GPS data from 58 minutes to 85 minutes. Figure 12 shows that the peak memory utilization is about 5.7 GB. The entire workflow takes 3 hours and 45 minutes to run, while the Stay Integration container requires the largest amount of computation time (98 minutes) among all the containers.

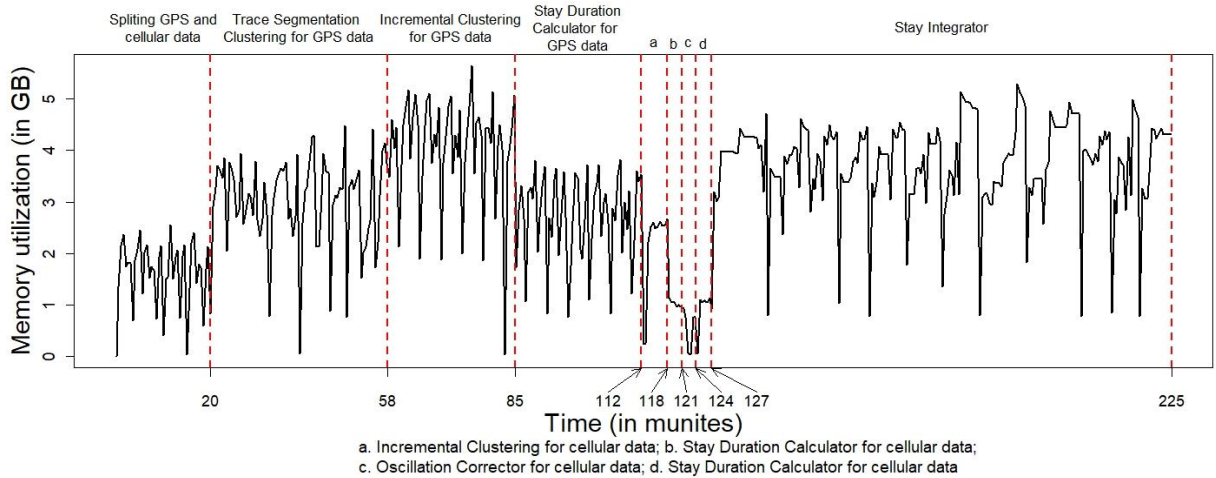

**Figure 12**. Computation time and memory utilization of the test workflow with 10 GB data comprising 60,000 users with data in March, April and November of 2019. The starting time is set to 0. Each red dashed line marks the time when the execution of a container is finished.

To examine how the computation time scales relative to the volume of data, Figure 13 shows the computation time of applying MAW for analyzing dataset of different sizes on the same type of Hyak computing node (Intel Xeon Gold 6230 processor and 64 GB memory). An approximately linear relationship between the data volume and the computation time is observed. These findings suggest that the developed MAW has manageable computational costs, in terms of both memory utilization and computation time, for analyzing large volumes of mobile data. This enhances the accessibility and interoperability of MAW on machines with limited computational resources (e.g. on one's personal computer), and the reusability of MAW to real-world cases in large scales where terabytes of mobile data is generated on a daily basis.

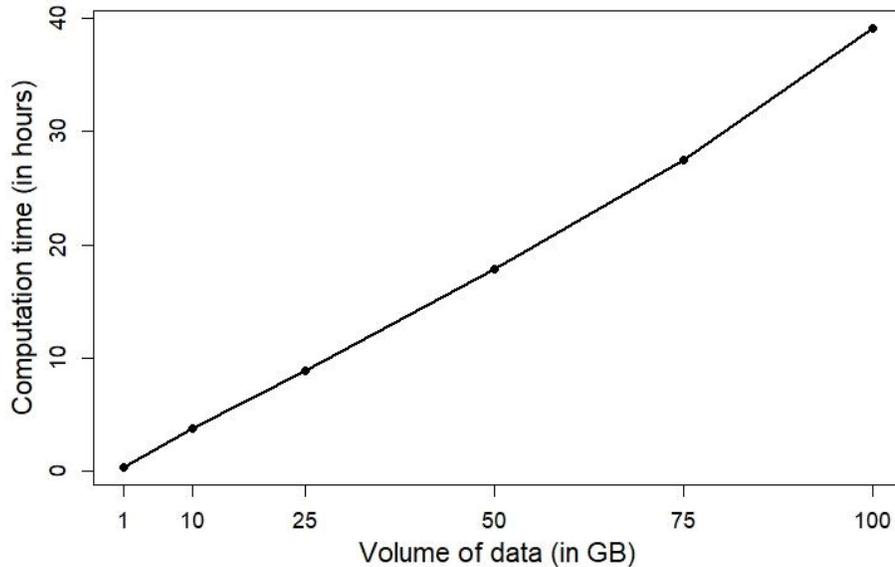

**Figure 13**. Time cost of running the test workflow for data of different volumes as input.

## 6. Concluding Discussions

This study has two unique contributions. First, it addresses the issues of accessibility, interoperability, reproducibility and reusability long existing in the human mobility research community. An important charge of human mobility research is to discover general truths about how individuals conduct their activity and travel patterns in time and space (Chen et al., 2016). This means that findings from one study shall be replicable in another study, with different datasets but the same set of analysis methods (National Academies of Sciences, Engineering, and Medicine, 2019). Replicability requires that analysis methods and codes developed to be accessible to others; the same set of results can be reproduced when the same methods are applied to the same data; and the software codes developed to implement some analysis methods can be reused for additional cases and/or different contexts for consistency checks. It is those considerations that initially motivated this study.

The present study is also motivated by that the results of mobility analysis form the very basis of transportation policies and investments that affect millions of people's lives. Though there has not been disturbing news about scientific fraud, recent years have witnessed a pattern of failing to replicate sometimes sound results across domains including psychology, economics, environmental, and medical sciences (Aarts et al., 2015; Camerer et al., 2016; Pexma and Lupker, 1995; Stupple et al., 2019). These disturbing trends, some of which are high-profile, have prompted multiple federal agencies including the National Science Foundation, the National Institutes of Health, and the National Academies of Sciences, Engineering, and Medicine to issue guidelines and reports on issues relating to reproducibility and replicability.

Compared to other fields, transportation research and in particular travel behavior and mobility research, lags behind in thinking and writing about, discussing, and integrating reproducibility and replicability into research. A search in the literature using keywords such as transportation, reproducibility, and replicability returns few studies. While there may be legitimate factors that may inhabit reproducibility and replicability (e.g., diverse methods and diverse data being used, confidentiality), one can also argue that the use of diverse methods and data are precisely the reasons for increasing reproducibility and replicability across studies. Figure 1 illustrates this need amid the larger background that more of these big datasets will be increasingly used for transportation investment and policy decisions. The fact that few to no study results from core transportation journals may also reflect the reluctance of leading journals to view such work as science and thus indirectly play a role that discourages the promotion of reproducibility and replicability in transportation research. The authors argue that it is time to engage transportation researchers to start conversations about these issues so that definitions and meanings of reproducibility and replicability can be better understood and severe, irreversible mistakes can be avoided. This need is especially dire for mobility analysis.

The diversity of mobility analysis methods and data also highlights the second contribution of this study: to advance our understanding of how different mobility analysis methods and their parameter settings affect the inferred mobility patterns. Since the debut of human mobility analysis using big mobile data in early 2000s, the collection of mobility analysis methods and algorithms has significantly expanded. And yet, comparative analysis of different mobility analysis methods is still few. This further limits the reusability of mobility analysis methods, as whether there are alternatives to the methods and how their parameter values should be reset to fit in a different context cannot be ascertained.

This study develops a mobility analysis workflow (MAW) so that the codes developed in this analysis can be accessible to, reproduced, and reused by others. Reusability supports replicability because a developed workflow for one mobility analysis can be reused on new datasets for consistency checks. As noted earlier, among the vast number of studies using mobile sensor data, the underlying data pre-processing procedures are often not reported and methods used to analyze the data and derive trajectories are many but there are few benchmarking studies comparing the results of different methods on the same dataset, or the same method on different datasets. This study represents an effort on the authors that puts a set of codes for mobility analysis into containers and workflows so that others can reproduce and replicate the current study and reuse them for their own studies.

Moreover, built upon MAW, this study tests and analyzes how different pre-processing and analysis methods affect inferred mobility patterns. Different workflows that incorporate different mobility analysis methods, different orders in applying the methods and different parameter value settings are designed and tested using MAW. The number of trips, radius of gyration (i.e. spatial distribution of trips) and departure time distribution (i.e. temporal distribution of trips) are compared among the results from different workflows. The results confirm the impacts of data pre-processing and analysis algorithms on the derived mobility patterns. These results are valuable information to researcher and practitioners in selecting appropriate mobility analysis methods and developing new ones, and enhance the reusability of mobility analysis methods in different contexts.

## Acknowledgements

Guan, Ren, and Chen are grateful to the funding provided by the US Federal Highway Administration (FHWA), the Washington State Department of Transportation (WSDOT), the US National Institute of Health (NIH) (1R01GM108731-01A1) and the Center for Teaching Old Models New Tricks (TOMNET), a University Transportation Center sponsored by the US Department of Transportation through Grant No. 69A3551747116. Ren was also supported by the University of Washington CoMotion Step Fund. Hung, Lloyd and Yeung are supported by NIH grant R01GM126019. Lloyd is also supported by the NSF Advanced Cyberinfrastructure Research Program (OAC-1849970). All authors would also like to thank Amazon Web Services for AWS Cloud Credits awarded for research. Authors also thank for the help provided by Ms. Grace Jia, a PhD student in the Department of Civil and Environmental Engineering at the University of Washington, who helped improve the readability of the tutorial. All views and errors are in the responsibility of the authors.

Hung and Yeung have equity interest in and employed by Biodepot LLC, which receives compensation from NCI SBIR contract number 75N91020C00009. The terms of this arrangement have been reviewed and approved by the University of Washington in accordance with its policies governing outside work and financial conflicts of interest in research.